\documentstyle[twoside,fleqn,espcrc2,epsf,bezier]{article}

\title{Phase Structure of Lattice QCD with Wilson fermion
at Finite Temperature} 

\author{S. Aoki\address{Institute of Physics, University of Tsukuba,
Tsukuba, Ibaraki 305, Japan}}

\begin{document}  

\begin{abstract}
We review our attempt at understanding the phase structure
of lattice QCD with the Wilson fermion formulation at finite temperature,
based on a spontaneous breakdown of parity-flavor symmetry.  
Numerical results demonstrating explicitly the spontaneous
breakdown of parity-flavor symmetry beyond the critical line at both zero 
and finite temperatures are reported.
The phase structure and order of the chiral transition are reported for
$N_f=2$,3, and 4 flavors, and the approach towards the continuum limit
is discussed.
\end{abstract}

\maketitle
\section{Introduction}

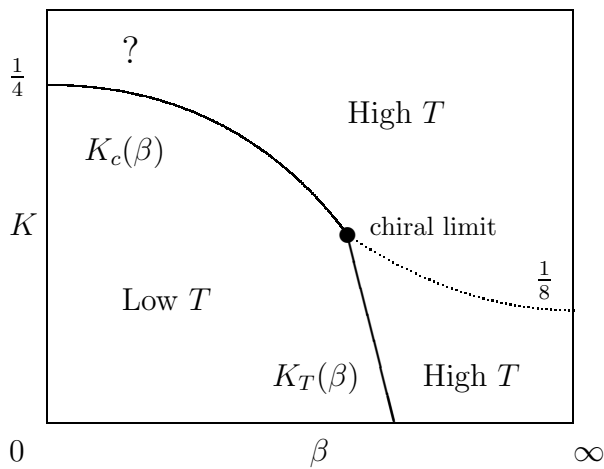
\begin{figure}
\begin{center}
\vspace*{-5mm}
\setlength{\unitlength}{1mm}
\begin{picture}(75,60)
\put(5,5){\line(1,0){70}}
\put(5,5){\line(0,1){55}}
\put(75,5){\line(0,1){55}}
\put(5,60){\line(1,0){70}}
\bezier{400}(5,50)(30,50)(45,30)
\put(45,30){\thicklines\line(1,-4){6.25}}
\put(45,30){\circle*{2}}
\bezier{50}(45,30)(60,20)(75,20)
\put(0,0){\large 0}
\put(40,0){\large $\beta$}
\put(0,30){\large $K$}
\put(75,0){\large $\infty$}
\put(0,50){\large $\frac{1}{4}$}
\put(70,23){\large $\frac{1}{8}$}

\put(48,30){chiral limit}
\put(15,20){\large Low $T$ }
\put(55,10){\large High $T$ }
\put(45,45){\large High $T$ }
\put(10,40){\large $K_c(\beta)$}
\put(35,10){\large $K_T(\beta)$}
\put(15,53){\Large  ?}
\end{picture}

\end{center}
\vspace*{-10mm}
\caption{A phase structure at finite temperature.}
\label{naive}
\end{figure}

Studying the finite-temperature QCD phase transition via the lattice 
regularization with the Wilson fermion formulation is known to be more 
difficult than with the staggered fermion formulation. This is because of
absence of an order parameter due to explicit chiral symmetry breaking 
introduced by the Wilson term.
A naive expectation for the phase structure in a $\beta = 6/ g^2$ 
vs. $ K = 1/(2 m_q a + 8)$  plane at finite temperature 
with fixed temporal lattice size $N_T$, which was the basis of early studies
of the phase transition\cite{earlywork}, is sketched in Fig.~\ref{naive}.
At zero temperature the pion mass vanishes along a line $K_c(\beta )$, the
critical line, which runs from $K_c(\beta=0)=1/4$ in the strong coupling limit
to $K_c(\beta=\infty)=1/8$ in the weak-coupling limit.
At finite temperature, which is realized at fixed $N_T \ll L$ where $L$ is a 
spatial lattice size, another critical line $K_T(\beta)$, 
the finite temperature phase transition line, should appear starting from
$K=0$, which corresponds to the deconfinement transition of pure gauge theory.
The critical line, though rigorously speaking some part of the line may 
merely show cross-over phenomena, is expected to extend into the non-zero $K$
region from below and hopefully to merge with the critical line $K_c$ at some 
$\beta$. The crossing point, if it exists, may be identified as the point
 of chiral phase transition, since at this point
the finite temperature phase transition occurs in the presence of 
a massless pion, or equivalently massless quarks.

In Fig.~\ref{naive} the region on the left of $K_T$ and below $K_c$
is in the low temperature phase,
while the region on the right of $K_T$ is in the high temperature phase.
The critical line $K_c$ in the high temperature phase is no more critical,
since the pion mass does not vanish in the high temperature phase,
as naively expected and confirmed by numerical simulations.
Therefore in Fig.~\ref{naive} a dotted line is assigned to the part
of the $K_c$ line 
included in a high temperature phase, where the pion mass does not vanish.

Although the phase structure above seems plausible at first
sight, several questions arise.
First of all, what is the meaning of $K_c$ at finite temperature ?
We define $K_c$ as a line of vanishing pion mass at zero temperature,
but according to the phase structure above the pion does not become massless
on $K_c$ in the high temperature phase. This suggests that 
a location of the line $K_c$ may depend on temperature if
it is defined as a line of vanishing pion mass at corresponding temperature.
A related question is the nature of the region above $K_c$ at small $\beta$, 
where a question mark is assigned in Fig.~\ref{naive}.
This region seems to belong to the low temperature phase since the pion
can become arbitrarily light in the region near $K_c$. 
If this is the case,
the line $K_T$ should continue to exist beyond the chiral transition point
in larger $K$ region, so that the low temperature phase above the $K_c$
can be separated from the high temperature phase.
This should be numerically confirmed.
A more fundamental question is how one can understand the existence of a
massless pion at zero temperature without recourse to chiral symmetry.

In this talk we try to answer these questions, based on interpretation
of the massless pion in the Wilson fermion formulation, using 
both analytic methods and numerical simulations.

\section{Phase structure at zero temperature}

\subsection{Parity-flavor breaking scenario}
In this subsection we briefly review our understanding of massless 
pions appearing at $K_c$ in the Wilson fermion formulation
at zero temperature based on refs.\cite{aoki1,aoki2,aoki3}.

Let us consider 2-flavor QCD.
Since pions become massless at $K=K_c$ we expect the following behavior
of the 2 point function for the corresponding operators
$\pi^a = \bar\psi i\gamma_5\tau^a\psi$ ($a=1,2,3$):
\begin{eqnarray*}
&\displaystyle\lim_{t\rightarrow \infty}& \frac{1}{V}
\sum_x \langle \pi^a (0,0) \pi^a (x,t) \rangle  \\
& = & \left\{
\begin{array}{ll}
Z_1 \exp [-t/\xi] , & K < K_c \\
Z_2 t^{-\alpha} , & K=K_c \\
\langle \pi^a  \rangle^2 + Z_3\exp[-t/\xi] , & K > K_c
\end{array}
\right. 
\end{eqnarray*}
where the correlation length $\xi$ is related to the mass of the $\pi$ meson
as $\xi = 1/ (m_\pi a)$, $\alpha$ is some exponent, and $ V=\sum_x$ is the 
space volume. 

Since an existence of massless $\pi^a$ meson implies a divergent correlation 
length at $K_c$, the operator corresponding to the $\pi^a$ meson
is expected to develop a long-range order for $K > K_c$ so that
$ \langle \pi^a \rangle \not=0$. This non-zero condensation
is possible for all $a$ in general, but using flavor symmetry
we can make a change of variables such that only one $\pi^a$ develops
a non-zero vacuum expectation value.
Hereafter, without loss of generality,
we assume that $\langle \pi^3 \rangle \not=0$
and $\langle \pi^{1,2} \rangle =0$.

This condensation spontaneously breaks
both parity and flavor symmetry, and a second order phase transition from 
the parity-flavor symmetric phase to
a parity-flavor broken one occurs at $K_c$.
Therefore a massless (neutral) $\pi$ meson 
corresponding to the $\pi^3$ operator
can be interpretated as a massless mode
associated with this second order phase transition,
not a Nambu-Goldstone boson. 
This neutral $\pi$ meson becomes massless only at the transition point $K=K_c$,
while the charged $\pi$ mesons corresponding to the linear combinations of
$\pi^1$ and $\pi^2$ operators stay massless in the broken phase
since they are the Nambu-Goldstone bosons associated with the
{\it flavor} symmetry breaking.

According to the above interpretation we expect near $K_c$,
$
(m_\pi a)^2 \sim (K_c-K)^{2\nu} \propto (M-M_c)^{2\nu}
$
with a critical exponent $\nu$, where we define $M=1/(2K)$ for later uses.
Since low energy properties of $\pi$ mesons may be described by an
effective 4-dimensional scalar field theory,
we can expect
the phase transition to be mean-field like up to logarithmic corrections
and therefore $\nu =1/2$, which reproduces the well-known PCAC relation 
$
(m_\pi a)^2 \propto m_qa
$
where quark mass  may be defined as $m_q a \simeq M-M_c$.

On the other hand, if we define the current quark mass via the axial 
Ward identities\cite{qmass}:
\begin{equation}
2 m_q^{WI} = \frac{\Delta_x \langle 0 \vert \bar\psi i\gamma_5\gamma_0 \psi (x)
\vert \pi \rangle}{\langle 0 \vert \bar\psi i\gamma_5\psi (x)\vert \pi 
\rangle} ,
\label{ward}
\end{equation}
this $m_q^{WI}$ automatically satisfies the PCAC relation 
$ m_\pi^2 \propto m_q^{WI} $ at small $m_\pi^2$.
However, since this $m_q^{WI}$ is not a tunable parameter, 
an existence of the {\it chiral} limit where $m_\pi$ becomes zero
is not ensured. Our scenario explains why such a tunning is possible.
Note that $m_q^{WI} \propto (M-M_c)$ if $\nu=1/2$.

This new scenario for the existence of the massless pion in the Wilson fermion
formulation has been confirmed analytically in the strong coupling and $1/N$
expansions. See refs.\cite{aoki1,aoki2} for details.

\begin{figure}[bt]
\centerline{\epsfxsize=65mm \epsfbox{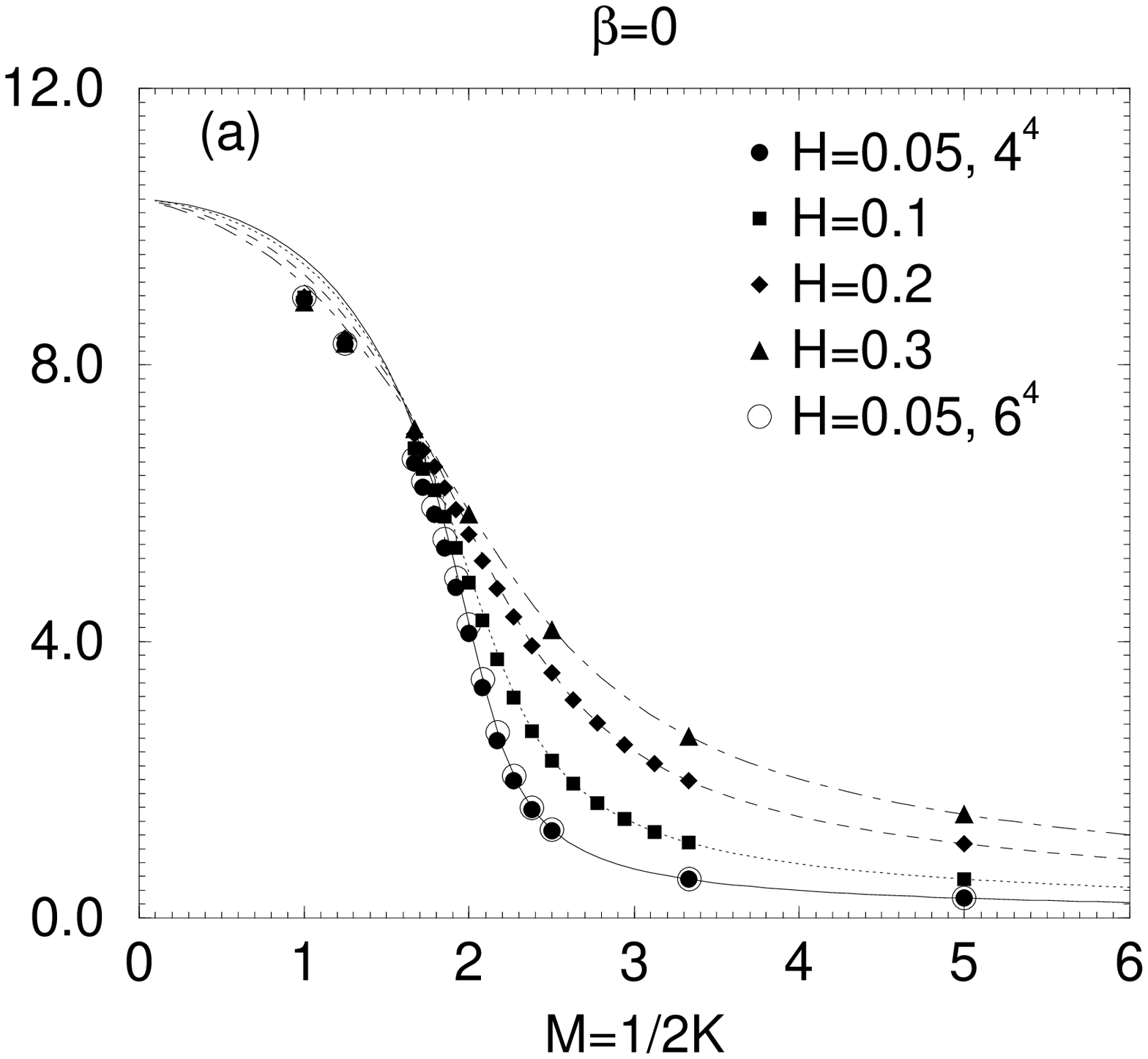}}
\centerline{\epsfxsize=65mm \epsfbox{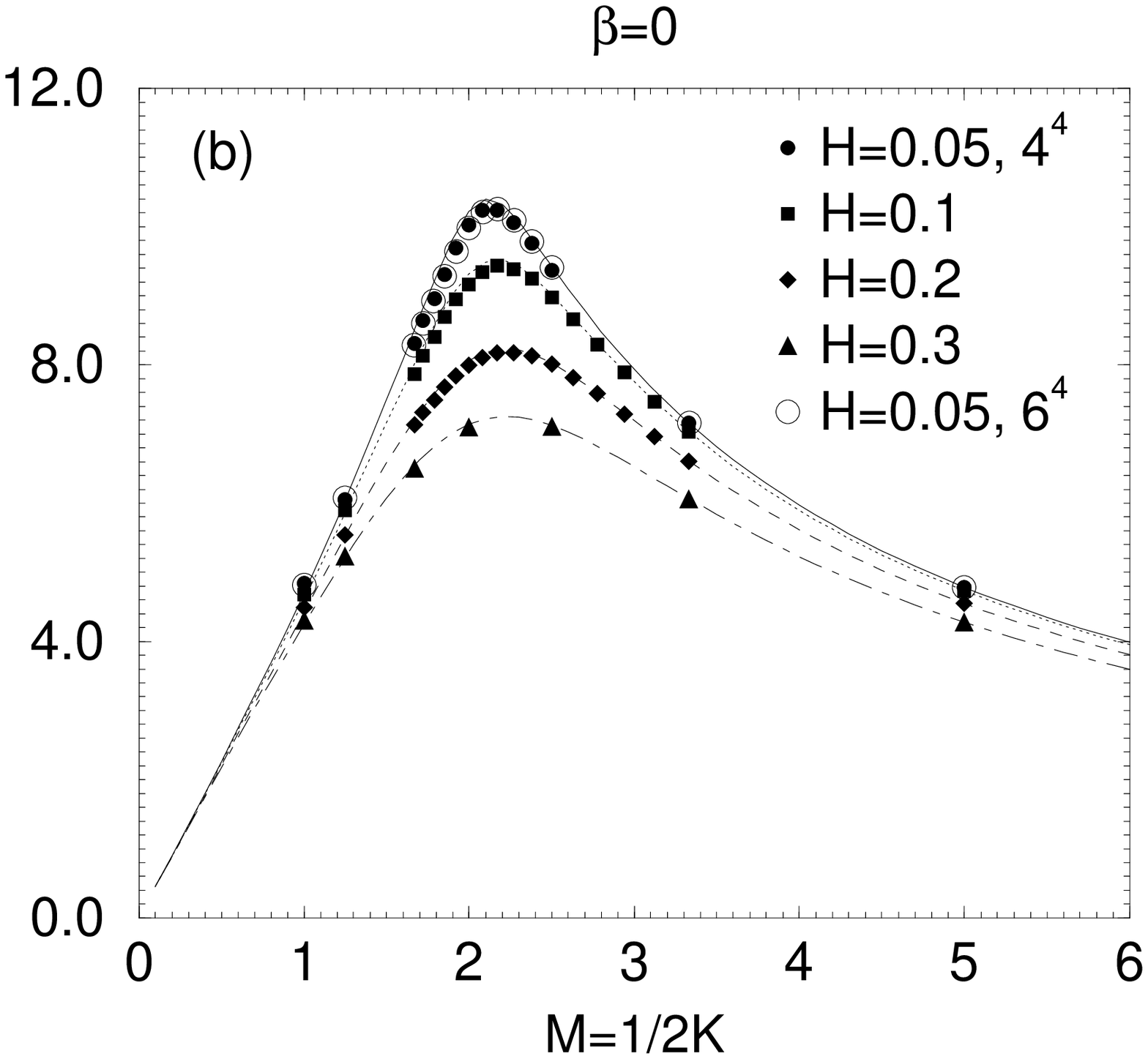}}
\vspace*{-10mm}
\caption{(a) $\langle \bar\psi i\gamma_5\tau^3\psi\rangle$
as a function of $M=1/2K$ at $\beta =0$ on a $4^4$ lattice
with $H=0.05$, 0.1, 0.2, 0.3 (solid symbols) and on a $6^4$ lattice
with $H=0.05$ (open circles). The lines are analytical predictions obtained at
$\beta =0$ in the large $N_c$ limit with the corresponding $H$.
(b) same for $\langle \bar\psi \psi\rangle$ .}
\label{ordb0}
\end{figure}

\subsection{Numerical evidence for the parity-flavor breaking phase}
\label{sec:b0}

In this subsection numerical evidence will be given for
the existence of the parity-flavor breaking phase in full QCD
with Wilson fermions.

In order to examine if $\pi^3 =i \bar \psi \gamma_5 \tau^3 \psi $
can develop a non-zero expectation value, we have to add a source term
$ i H \bar \psi \gamma_5 \tau^3 \psi $ to the Wilson fermion action.
The reason for this is as follows.
Since 
$
{\rm Re} [  i \bar \psi \gamma_5 \tau^3 \psi ]
$ 
changes its sign according to the sign of $H$ for
{\it an arbitrary}  configuration, it is easy to show,
at $H=0$ on a finite lattice, that

\[
{\rm Re} [ i \bar \psi \gamma_5 \tau^3 \psi ] = 0
\]
{\it without} averaging over configurations.
Therefore one can not see any signal of a non-zero value of
$\langle i \bar \psi \gamma_5 \tau^3 \psi \rangle $ 
from simulations with $H=0$.
This is in contrast to the case of
the magnetization of the Ising model, where it takes non-zero values
on a given configuration for zero magnetic field
and may change its sign from configuration to configuration.
To obtain the correct value of 
$\langle i \bar \psi \gamma_5 \tau^3 \psi \rangle $
in the Wilson fermion formulation,
one has to take limits in the following order:
\[
\lim_{H\rightarrow 0} \lim_{V\rightarrow\infty}
\langle i \bar \psi \gamma_5 \tau^3 \psi \rangle .
\]
This means that one should make an extrapolation to $H=0$ from
a series of values at non-zero $H$'s 
on finite but sufficiently large lattices.

Fortunately, even in the presence of the source term,
the Wilson fermion determinant is still real positive, so that
the hybrid Monte Carlo method can be utilized.

In order to establish the parity-flavor breaking phase
we have carried out a full QCD simulation with $N_f = 2$ at $\beta =0$,
where analytic expressions for the order parameters are available
for non-zero external field $H$ within $1/N_c$ expansion\cite{aoki1}. 

In Fig.~\ref{ordb0}, $\langle \bar\psi i\gamma_5\tau^3 \psi \rangle$
and 
$\langle \bar\psi \psi \rangle$  are plotted as a function of $M =1/ 2 K $
at $\beta =0$ on a $4^4$ lattice with $H= 0.05,0.1,0.2,0.3$ and
on a $6^4$ lattice with $H=0.05$,
together with the analytic predictions with the corresponding $H$
in the strong coupling and large $N_c$ limits.
Since the agreement between the numerical data and the analytic predictions
is very good, it can be established at least in the strong coupling limit 
that the parity and flavor symmetry breaking phase transition really occurs
at $M=M_c=2$ ($K=K_c=1/4$) and $\langle \bar\psi i\gamma_5\tau^3 \psi \rangle$
is non-zero in the $H\rightarrow 0$ limit at $M<M_c$ ($K>K_c$).
We have also shown that finite size effects on the order parameter
is very small by comparing data on a $4^4$ lattice with those on
a $6^4$ lattice or with the analytic predictions which correspond to
the values in the infinite volume limit.

\begin{figure}[tb]
\vspace*{-10mm}
\centerline{\epsfxsize=70mm \epsfbox{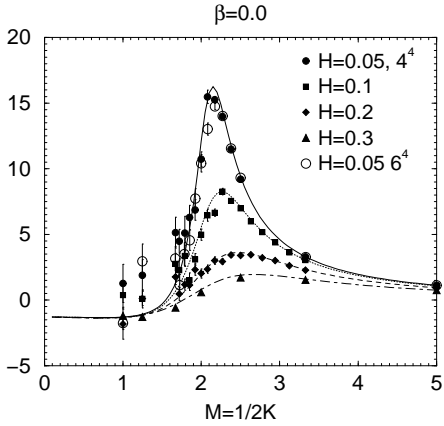}}
\vspace*{-10mm}
\caption{Susceptibility $\chi_{55}$
as a function of $M=1/2K$ at $\beta =0$ on a $4^4$ lattice
(solid symbols) and on a $6^4$ lattice 
(open circles). 
}
\label{susp}
\end{figure}

\begin{figure}[tbh]
\vspace*{-10mm}
\centerline{\epsfxsize=70mm \epsfbox{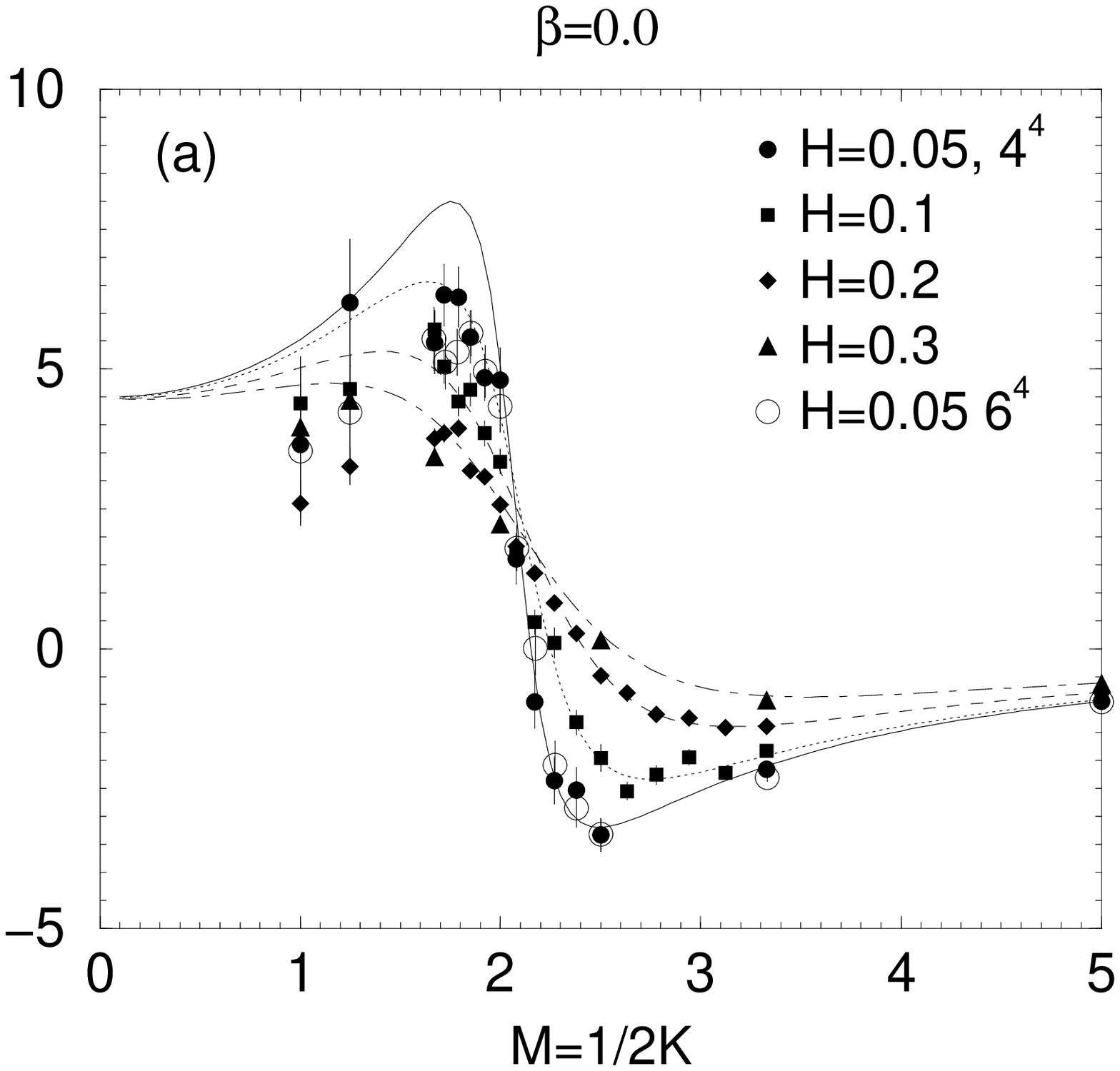}}
\centerline{\epsfxsize=70mm \epsfbox{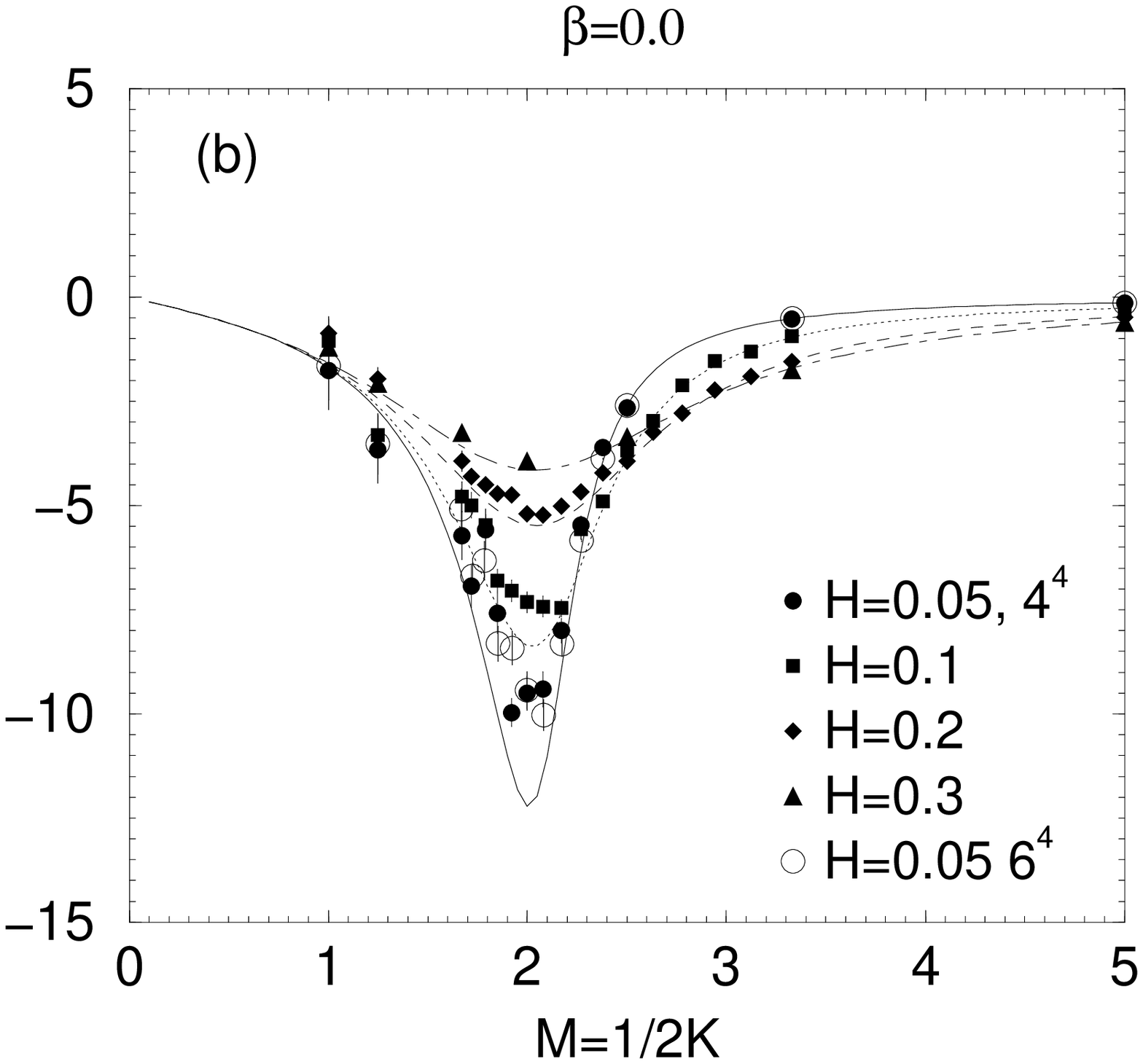}}
\vspace*{-10mm}
\caption{(a) $\chi_{00}$. (b) $\chi_{50}$.}
\label{susp2}
\end{figure}

As a non-trivial check we have calculated susceptibilities $\chi_{55}$,
$\chi_{00}$, and $\chi_{50}$, which are defined by
\begin{eqnarray*}
\chi_{55} &=& \frac{1}{V}\left\langle \left[
\sum_x \bar\psi i\gamma_5\tau^3 \psi (x)\right]^2 
\right\rangle \\
\chi_{00} &=& \frac{1}{V}\left\langle \left[\sum_x \bar\psi \psi (x)
\right]^2 \right\rangle \\
\chi_{50} &=& \frac{1}{V}\left\langle \sum_x \bar\psi 
i\gamma_5\tau^3 \psi (x)
\times \sum_y \bar\psi \psi (y) \right\rangle .
\end{eqnarray*}
We have plotted $\chi_{55}$ in Fig.~\ref{susp}
and $\chi_{00}$, $\chi_{50}$ in Fig.~\ref{susp2},
together with the analytic predictions.
The agreement between the numerical data and the analytic predictions
is reasonably good, except for deviations
observed deep in the broken phase,
where our analytic formula becomes inaccurate in the presence of massless
Nambu-Goldstone charged pions.
These results for the susceptibilities also support our 
conclusion about the existence of the parity-flavor breaking phase.

\subsection{Phase structure at small $\beta$}

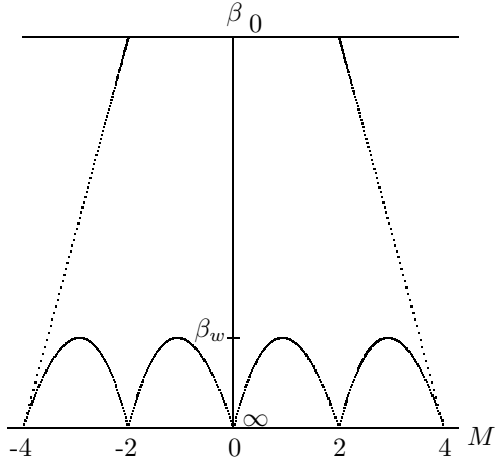
\begin{figure}[bt]

\begin{center}
\vspace*{-10mm}
\setlength{\unitlength}{0.4mm}
\begin{picture}(150,150)
\put(10,10){\line(1,0){150}}
\put(85,10){\line(0,1){130}}
\put(15,140){\line(1,0){145}}
\put(90,10){\makebox(5,5){$\infty$}}
\put(165,5){\makebox(5,5){$M$}}
\bezier{100}(15,10)(50,140)(50,140)
\bezier{100}(15,10)(35,70)(50,10)
\bezier{100}(50,10)(65,70)(85,10)
\bezier{100}(155,10)(120,140)(120,140)
\bezier{100}(155,10)(135,70)(120,10)
\bezier{100}(120,10)(100,70)(85,10)
\put(12,1){\makebox(5,5){-4}}
\put(47,1){\makebox(5,5){-2}}
\put(83,1){\makebox(5,5){0}}
\put(118,1){\makebox(5,5){2}}
\put(153,1){\makebox(5,5){4}}
\put(83,145){\makebox(5,5){$\beta$}}
\put(90,142){\makebox(5,5){0}}
\put(83,40){\line(1,0){4}}
\put(75,40){\makebox(5,5){$\beta_w$}}
\end{picture}

\end{center}
\vspace*{-10mm}
\caption{The expected phase structure of lattice QCD with Wilson fermion in
$\beta$ - $M$ plane.}
\label{phase}
\end{figure}

In the previous subsection we have numerically established that 
the parity-flavor breaking phase indeed exists in the strong coupling limit.
This phase should continue to exist in the strong coupling region,
since a strong-coupling expansion, which support an existence of 
the phase\cite{aoki1}, is convergent around $\beta=0$.
However an understanding of the phase structure
in the weak coupling region is particularly important,
since the continuum limit can be defined only in this region.
Therefore, in this subsection,
we consider how the phase structure changes
toward the weak coupling region.

\begin{figure}[tb]
\centerline{\epsfxsize=70mm \epsfbox{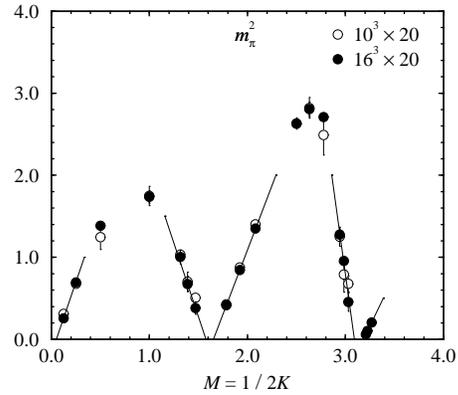}}
\vspace*{-10mm}
\caption{Pion mass squared $m_\pi^2$ at zero temperature as a function of
    $M=1/2K$ at $\beta=6.0$.  Open circles are for a $10^3\times20$ lattice,
    and filled circles for a $16^3\times20$ lattice.  Lines show linear fits of
    $m_\pi^2$ in $M$.}
\label{pmass6.0}
\end{figure}

In Fig.~\ref{phase} we propose the expected phase structure
of lattice QCD with 2 flavor Wilson fermions in the $\beta$ vs. $M=1/2K$ plane,
based on the reasoning mentioned later.
At $\beta < \beta_w$ there are only two critical lines, while
there are ten critical lines at $\beta > \beta_w$, and 5
points where two lines meet at $\beta =\infty$. 
At $M=4$ on can define the usual continuum limit 
where only a physical mode of the sixteen zero modes of the lattice fermion
becomes light, while the continuum limits at other 4 points
contains light doubler modes.
For example 
4 doubler modes, which has one of the momentum component near $\pi/a$,
appear as light fermions at $M=2$.
Note that the phase diagram is shown to be symmetric under the 
$M\rightarrow -M$ transformation.

There are several evidence for the existence of multiple 
critical lines 
in the weak coupling region. First of all, a phase structure
similar to Fig.~\ref{phase} has been found for the Gross-Neveu model
in 2 dimensions in the large $N$ limit, regulated on the lattice using
Wilson fermions\cite{aoki1,aoki3,EN}.
The existence of a multiple structure of the critical lines
has been also suggested from
studies of eigen-value spectra of the Wilson fermion matrix
on quenched configurations\cite{U1,SDB} and on
full QCD configurations\cite{BLLS}.

To establish the multiple structure of the critical lines
we performed new simulation using quenched configurations
{\it without} adding the external field $ i H \bar\psi \gamma_5\psi$.
We have
calculated the pion mass at $\beta=6.0$ on a $10^3\times20$ and a
$16^3\times20$ lattice to find the location of multiple $K_c$'s.
The pion propagator is averaged over $10$--$20$
configurations.

We plot our results for the pion mass squared in Fig.~\ref{pmass6.0} as a
function of $M=1/2K$. 
A reasonable agreement of results for $10^3$ and $16^3$ spatial sizes shows 
that finite spatial size effects are not severe in our data.  
We observe
that there exist four more critical values of vanishing pion mass in addition
to the conventional one located at the right most of the figure.  
The existence of 5 critical points for $M > 0$ at fixed $\beta$
supports the phase structure proposed in Fig.~\ref{phase}.

\section{Phase structure at finite temperature}

Since we now understand the phase structure of lattice QCD with the
Wilson fermion formulation at zero temperature,
we can extend our investigation to the case at finite temperature.

\subsection{Expected phase structure at fixed $N_T$}

\begin{figure}[bt]
\centerline{\epsfxsize=80mm \epsfbox{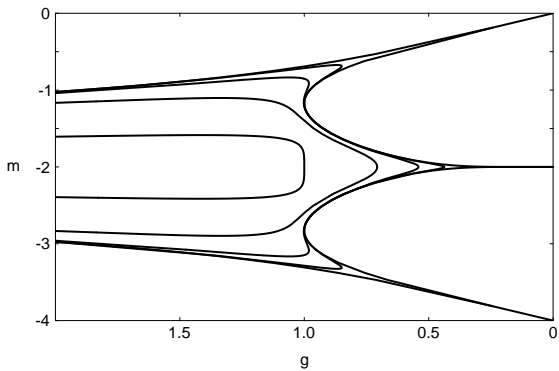}}
\vspace*{-10mm}
\caption{Critical line for the lattice Gross-Neveu model on $(g,m)$ plane. 
Temporal lattice size equals $N_t=\infty$, 16, 8, 4, and 2 from outside
to inside.}
\label{GN}
\end{figure}

In this subsection, based on the finding at zero temperature,
we propose the  phase structure for a fixed temporal
extension $N_T$ corresponding to finite temperature.

We first consider how the multiple structure of critical lines is modified
for a finite $N_T$. An important hint for the answer to this question
again come from the Gross-Neveu model in 2 dimensions. 
In Fig.~\ref{GN} we plot the critical line corresponding to $m_\pi =0$
on the $(g,m)$ plane for $N_T = \infty$,16,8,4 and 2 where $g$ and $m$
are the bare coupling and quark mass, respectively, the latter related to the 
hopping parameter through $K =1/(2m+4)$.
The crucial 
feature revealed in Fig.~\ref{GN} is that the three cusps of the critical line
retract from the weak coupling limit $g=0$ for finite temporal lattice sizes,
forming a single continuous line which shifts toward strong coupling as $N_T$
decreases. Thus, for a finite $N_T$, the critical line is absent for 
sufficiently weak coupling.

\begin{figure}[bt]

\begin{center}
\vspace*{-10mm}
\setlength{\unitlength}{1mm}
\begin{picture}(75,65)
\put(5,5){\line(1,0){70}}
\put(75,5){\line(0,1){60}}
\put(5,5){\line(0,1){60}}
\put(5,65){\line(1,0){70}}
\put(0,35){\large $K$}
\put(0,0){\large 0}
\put(40,0){\large $\beta$}

\put(5,57){\line(4,-1){70}}
\put(5,23){\line(1,0){70}}
\bezier{1000}(5,55)(65,40)(5,25)
\put(32,65){\thicklines\line (1,-5){12}}

\put(45,10){\Large $K_t$}
\put(50,32){\Large $K_c(T=0)$}
\put(55,30){\vector(0,-1){5}}
\put(55,37){\vector(0,1){5}}
\put(7,40){\Large $K_c(T\not= 0)$}
\put(15,37){\vector(0,-1){9}}
\put(10,10){\large Low T}
\put(10,60){\large Low T}
\put(60,60){\large High T}
\put(60,10){\large High T}

\end{picture}

\end{center}
\vspace*{-10mm}
\caption{The phase structure of lattice QCD expected for finite $N_T$.}
\label{phaseFT}
\end{figure}
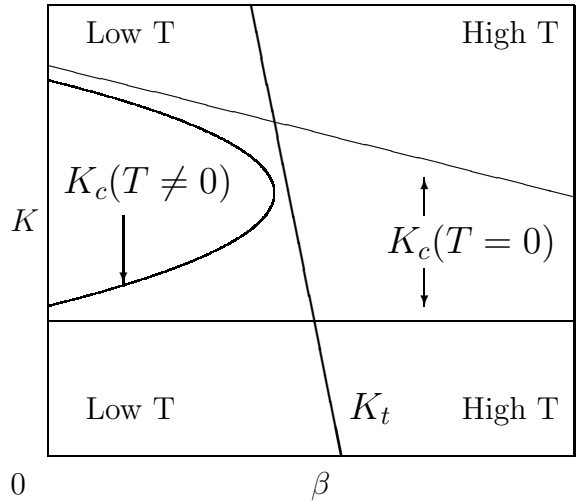

The close resemblance of the Gross-Neveu model and QCD such as the asymptotic 
freedom and the spontaneous chiral symmetry breaking
suggests that a similar behavior of the critical line will hold for QCD, 
except that the number of cusps should increase to five because of the 
difference in the number of dimensions, as
seen in the previous section. For the finite temperature case, an apparent
disappearance of the critical line toward weak coupling has been noticed in
previous studies\cite{milc,qcdpax1}. 
This means that $K_c$ at finite temperature differs from
$K_c$ at zero temperature,
if we define the $K_c$ as a line of vanishing pion (screening) mass 
at the corresponding temperature.

In addition to the deformation of 
the critical lines at finite temperature, the thermal transition line 
$K_t(\beta)$, which is absent in the Gross-Neveu model, should appear in QCD.
As discussed in the introduction, in order to
define the unique point of chiral phase transition at fixed $N_T$,
the thermal line $K_t(\beta)$ has to cross the critical line $K_c(\beta)$.
However the line $K_t$ can not go into the parity-flavor 
breaking phase, since the massless pions exist in
the phase so that it belongs to the low temperature phase.
Therefore the line $K_t$ never cross the line $K_c$ defined at fixed $N_T$.
In other words the region close to the critical line has to be
in the low temperature phase even after it turns back toward strong coupling.
This means that the thermal line should run past the turning point of the 
critical line\footnote{Logically it may be possible that the line $K_t$
touches $K_c (T\not= 0)$ without crossing it.}
and continue toward larger value of $K$, as sketched in 
Fig.~\ref{phaseFT}. Although $K_t$ seems to cross $K_c(T=0)$ as seen in 
Fig.~\ref{phaseFT}, no special criticality associated with massless quarks
occurs at the crossing point. Therefore, if the above phase structure is true,
we can not assign an unique point to define the point of chiral phase
transition.

\subsection{Results for $N_f$ = 2 at $N_T$=4}
\label{sec:nf2}

\begin{figure}[bt]
\centerline{\epsfxsize=70mm \epsfbox{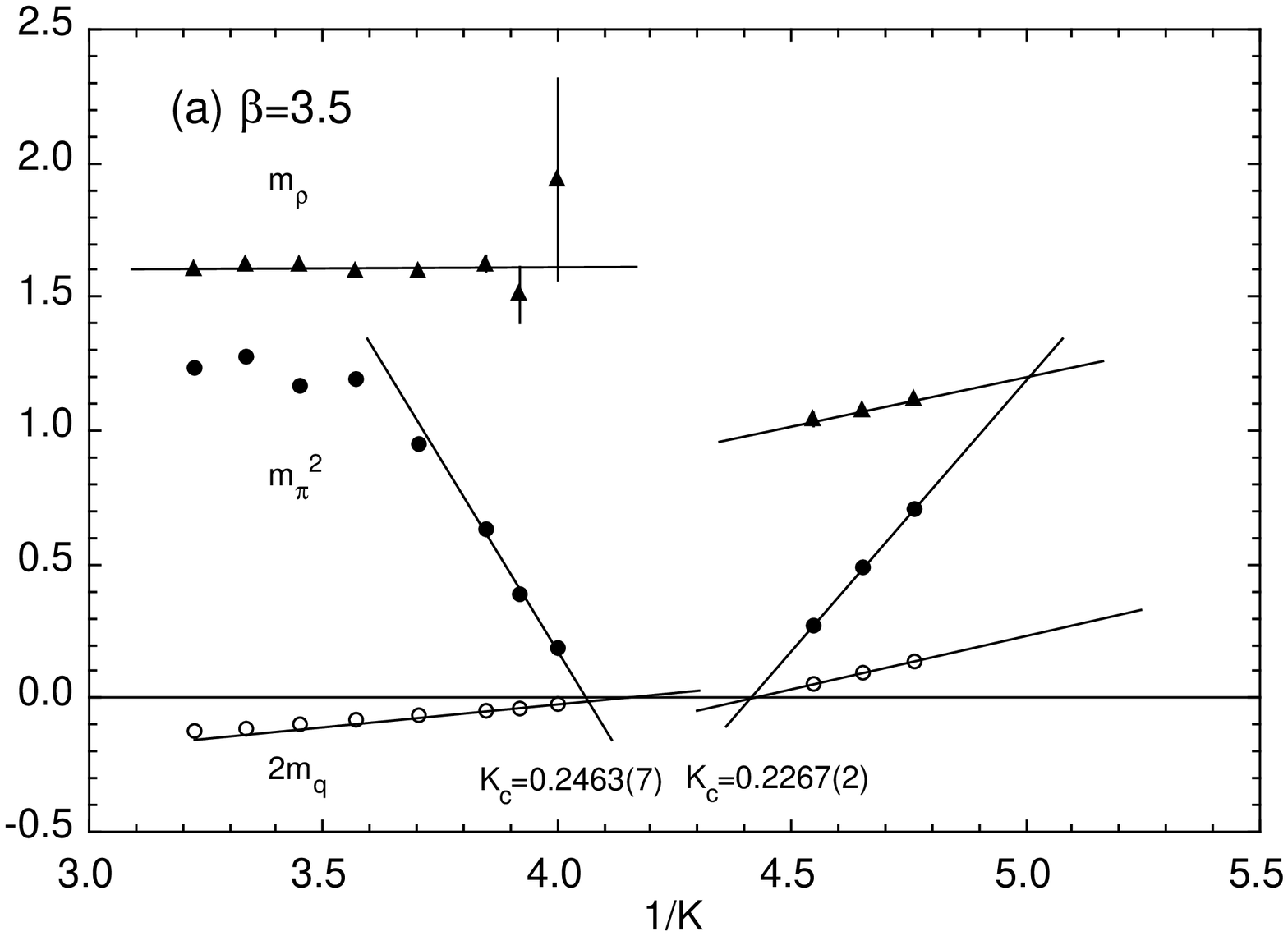}}
\centerline{\epsfxsize=70mm \epsfbox{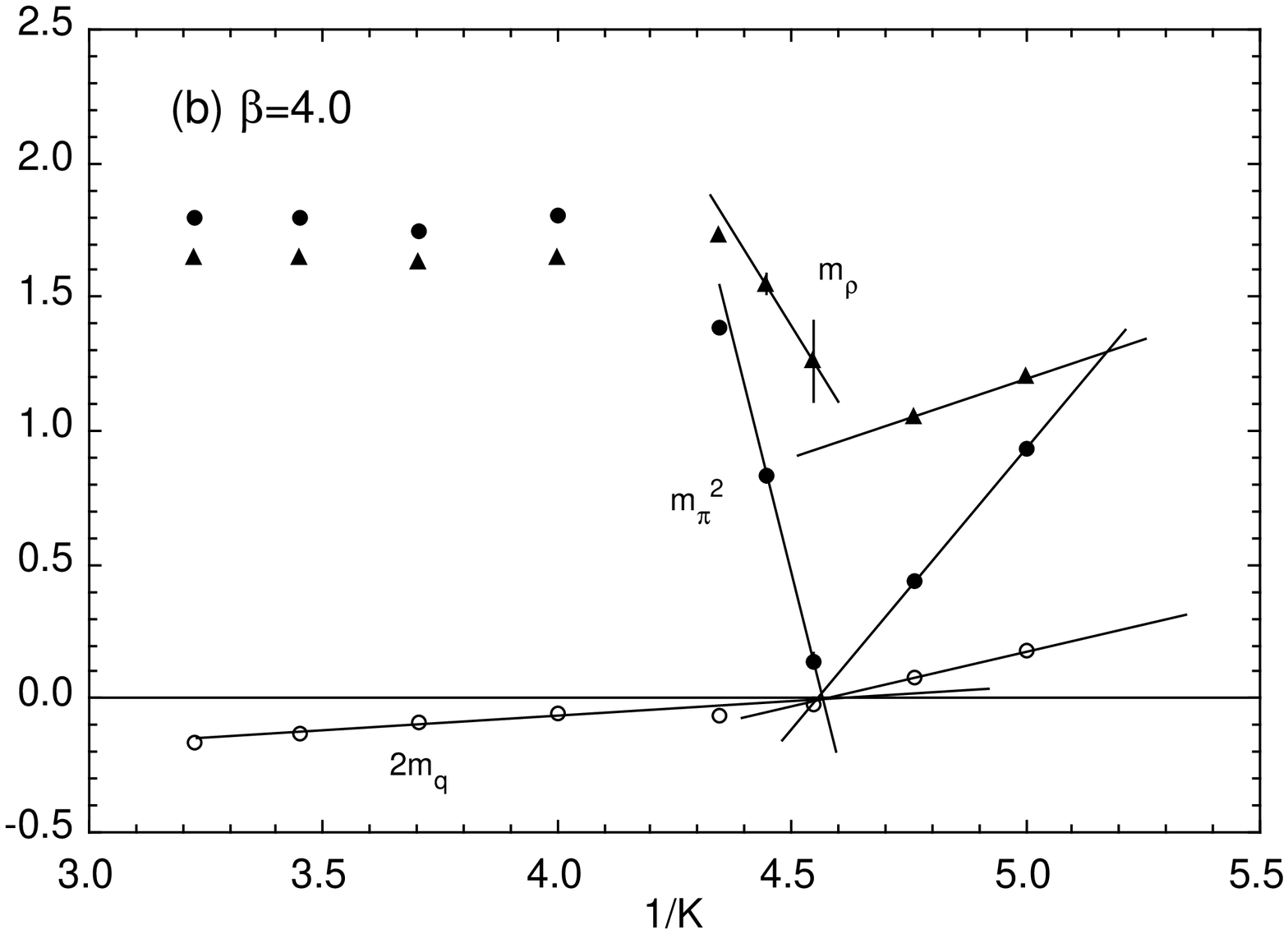}}
\vspace*{-10mm}
\caption{$\pi$ and $\rho$ screening masses and quark mass for $N_f=2$ full QCD
with Wilson quark action obtained on an $8^3\times 4$ lattice periodically
doubled in one of spatial directions. Lines are linear fits to three data 
points.}   
\label{mass_b35-40} 
\end{figure}

In this and the following subsections we review results supporting
the phase structure proposed in the previous subsection. 
Here we report results for the 2-flavor case
obtained for the temporal size $N_T=4$
with the standard one plaquette action for gauge fields.

We first show evidence that the cusps of the critical line 
retract from the weak coupling limit for a finite $N_T$ 
as speculated in the previous subsection.
For this purpose we have carried out hybrid Monte Carlo simulations
of $N_f=2$ full QCD using the Wilson quark action with $H=0$
on an $8^3\times 4$ lattice\cite{AUU1}.
In Fig.~\ref{mass_b35-40}(a) we present our results for the $\pi$ and $\rho$ 
screening masses, measured on the periodically doubled lattice
in the spatial directions $8^3\times 4\rightarrow (8\times2)\times 8^2
\times 4$, at $\beta =3.5$ as a function of $1/K$. Also plotted is the quark
mass defined via the axial Ward identity, eq.(\ref{ward}).

As usual the pion mass squared and the quark mass linearly vanish almost 
simultaneously at the conventional critical point $K_c\approx 0.2267$,
while another critical point at $K_c\approx 0.2454$
can be clearly observed from the behavior of the pion mass squared and 
the quark mass in the larger $K$ region.
The existence of two critical points close to each other is quite
analogous to Fig.~\ref{pmass6.0} in quenched QCD at zero temperature.
\footnote{Although we have not carried out further simulations at larger $K$
to identify 3 more critical points, there is no reason to doubt their
existence.}
On the other hand
Fig.~\ref{mass_b35-40}(b) shows how the behavior of these masses
changes at $\beta=4.0$.
We observe that the gap between the two critical values has either become 
extremely narrow or disappeared. 
These results at two $\beta$ values are consistent with the expected
deformation of the critical line at fixed $N_T$ as shown in 
Fig.~\ref{phaseFT}.

\begin{figure}[bt]
\centerline{\epsfxsize=70mm \epsfbox{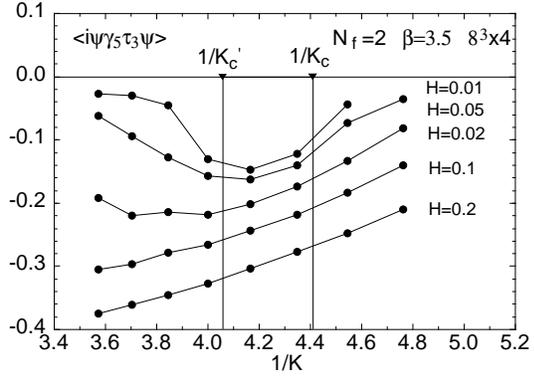}}
\vspace*{-10mm}
\caption{Parity-flavor order parameter as a function of $1/K$ for various
values of external field $H$ at $\beta=3.5$ on an $8^3\times 4$ lattice
for $N_f=2$.}  
\label{p5_b35}
\end{figure}

\begin{figure}[bt]
\centerline{\epsfxsize=70mm \epsfbox{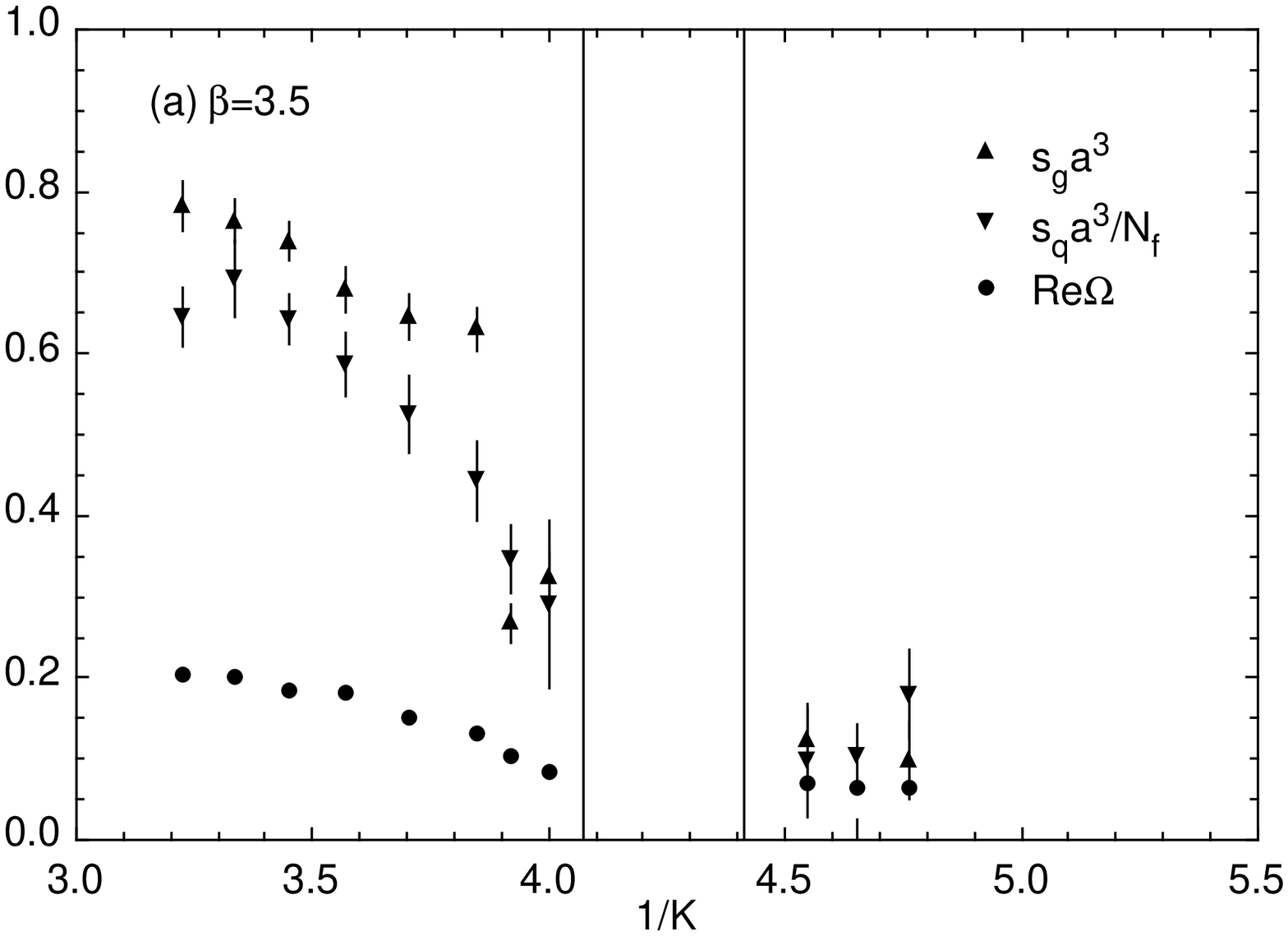}}
\centerline{\epsfxsize=70mm \epsfbox{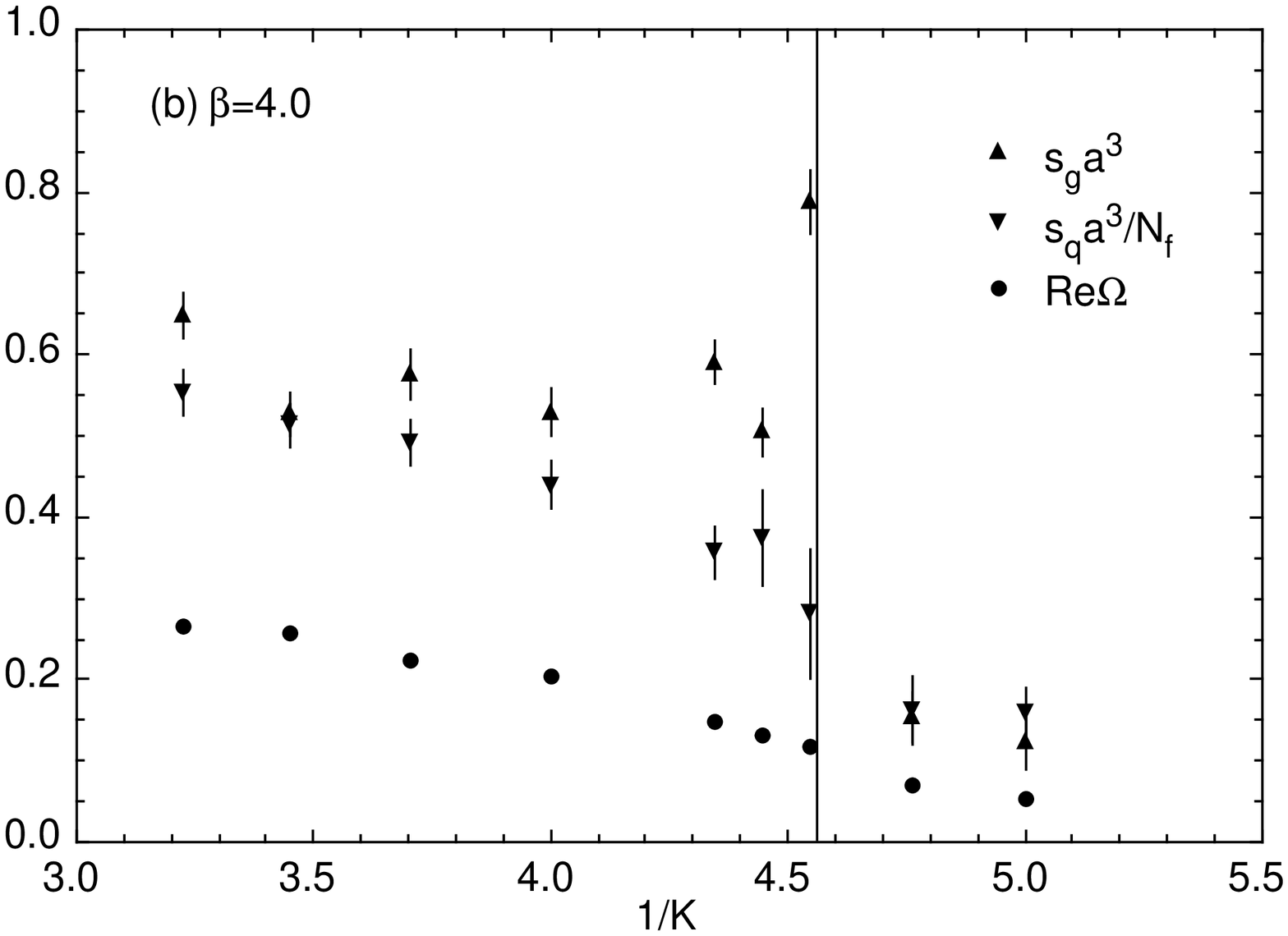}}
\vspace*{-10mm}
\caption{Real part of Polyakov line($Re\Omega$), gluon ($s_g$) and 
quark ($s_q$) entropy density in
lattice units for $N_f=2$ full QCD with Wilson quark action obtained on an
$8^3\times 4$ lattice. }    
\label{thermal_b35-40}
\end{figure}

Now one may ask whether the parity-flavor symmetry is indeed
spontaneously broken in the region between two  $K_c$'s at $\beta =3.5$.
In order to confirm this we have carried out new simulations
with non-zero $H$\cite{AUU2}.
In fig.~\ref{p5_b35} we plot the parity-flavor breaking order parameter 
$<\overline{\psi}i\gamma_5\tau_3\psi>$ as a function of $1/K$ for 
$H=0.2$,0.1,0.05,0.02 and 0.01. 
Vertical lines mark the position of the critical points at $\beta=3.5$.  We
clearly observe that the order parameter
between the two critical values $K_c$ and $K_c^\prime$ tends to a
non-vanishing value as the external field
$H$ is reduced, while it decreases toward zero outside,
supporting the spontaneous breakdown of parity-flavor symmetry 
between the two critical points.

\begin{figure}[bt]
\centerline{\epsfxsize=70mm \epsfbox{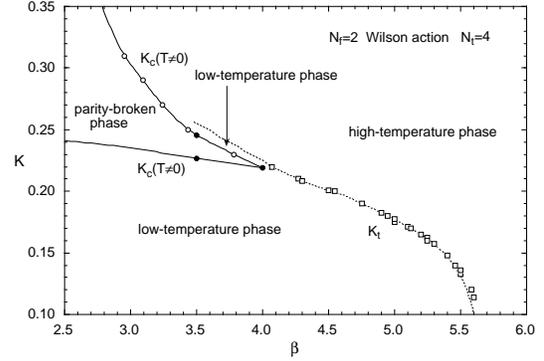}}
\vspace*{-10mm}
\caption{Phase structure of $N_f=2$ QCD with Wilson quark action.
Estimates of the critical line for $N_T=4$ (solid line) and the
thermal transition line (dotted line) are shown.}
\label{phaseNF2}
\end{figure}

We next examine if the thermal line $K_t$
run past the turning point of the critical line around $\beta =4.0$
and  continue towards larger values of $K$,
so that the region close to the critical line is in the low-temperature 
phase even after it turns back toward strong coupling. 
For that purpose we have measured thermodynamic observables,
the real part of the Polyakov line and quark and gluon entropy densities
in lattice units,
and plot them as a function of $1/K$ at $\beta =3.5$
in Fig.~\ref{thermal_b35-40}(a).
The two vertical lines again show the position of the critical points
estimated from the pion screening mass. 
Large values of these observables suggest that
the system is in the high temperature phase for smaller values of
$1/K$.
However they decrease as $1/K$ increases
toward the critical point at $1/K_c  \approx 4.075$, 
becoming roughly similar in magnitude to those on the other side of the 
conventional critical point at $1/ K_c\approx 4.411$.
Since the conventional critical point is in the low-temperature phase,
a similarity of these thermodynamic quantities in magnitude
support our expectation that the other critical point is also in
the low-temperature phase. 

Behaviors of thermodynamic observables 
at $\beta=4.0$ are shown in Fig.~\ref{thermal_b35-40}(b),
together with the vertical line which
marks the  point where the two linear extrapolations of the pion screening mass
squared in Fig.~\ref{mass_b35-40}(b) cross  each other.  
Since the thermodynamic quantities around the line stay small,
the region around the ``crossing point'' is likely to be also in the
low-temperature phase.
On the other hand
the increase of the three quantities across the
line suggests that the thermal line runs very
close to the critical line at $\beta=4.0$. 
Within the resolution of our numerical simulation, however,
we can not exclude the possibility that the thermal transition line touches
the critical line around $\beta=4.0$ and run away again for larger $K$.

Overall results in this subsection are consistent with our proposed
phase diagram in Fig.~\ref{phaseFT}. For definiteness, in 
Fig.~\ref{phaseNF2}, we draw the phase diagram of two flavor QCD
for $N_T =4$,
mainly based on our finding. We estimate the location of the critical line
$K_c$ (solid line) from
the points of vanishing pion screening mass (solid circles) as usual,
supplementing the points where the inverse of the number of conjugate gradient
iterations in the runs is linearly extrapolated to zero(open circles).
Here we assume that the turning point of the critical line
is located at $\beta =4.0$.
The dotted line represents the thermal transition line $K_t$ estimated
from previous works (open circles)\cite{earlywork,milc,qcdpax1}.
Our drawing of the dotted line continues above the upper part of the cusp of 
$K_c$,
though its precise location can not be determined from our numerical data.

\subsection{RG improved gauge action for $N_f$=2}
\subsubsection{Evidence of the cusp structure of the critical line}

\begin{figure}[bt]
\centerline{\epsfxsize=65mm \epsfbox{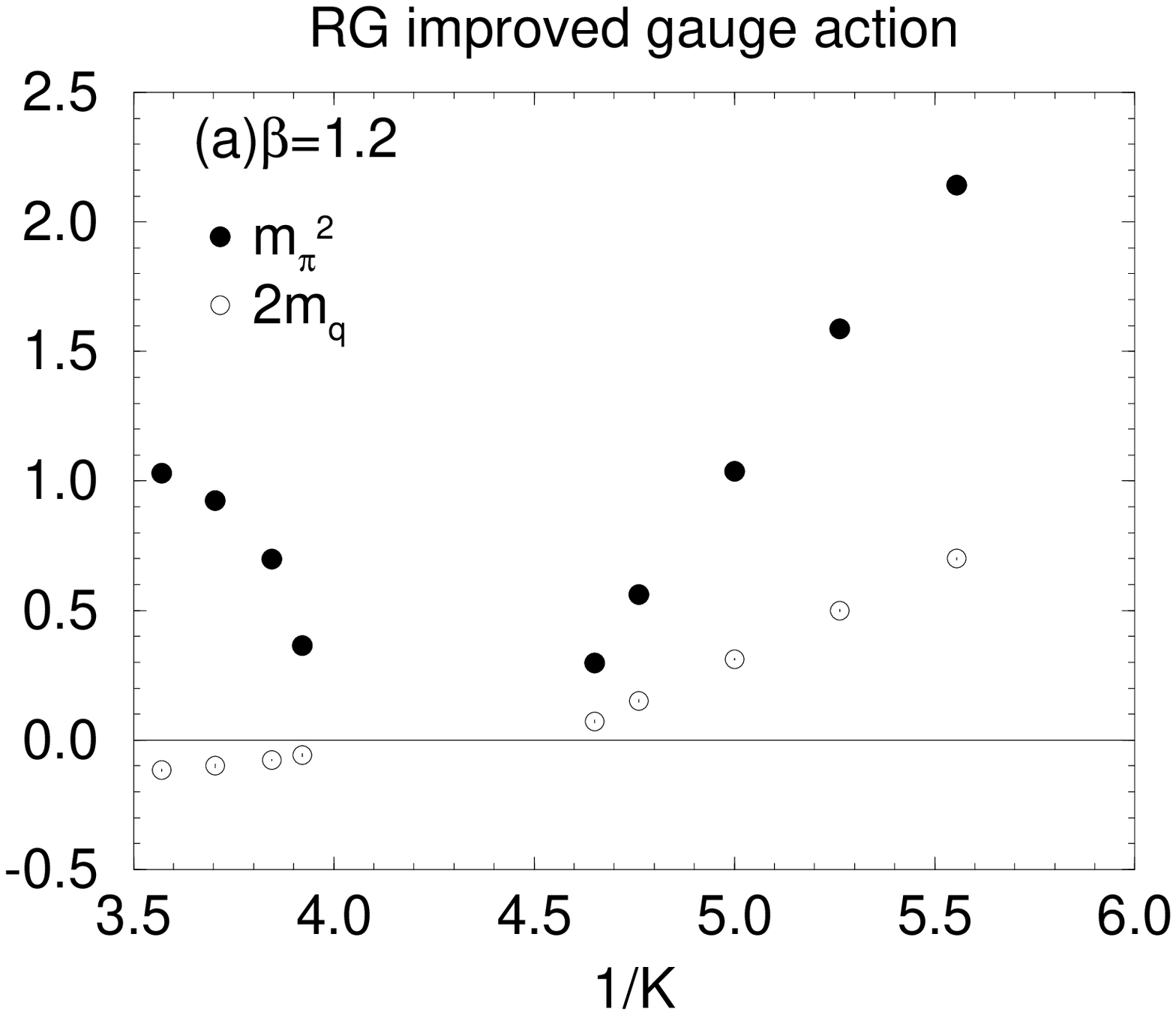}}
\centerline{\epsfxsize=65mm \epsfbox{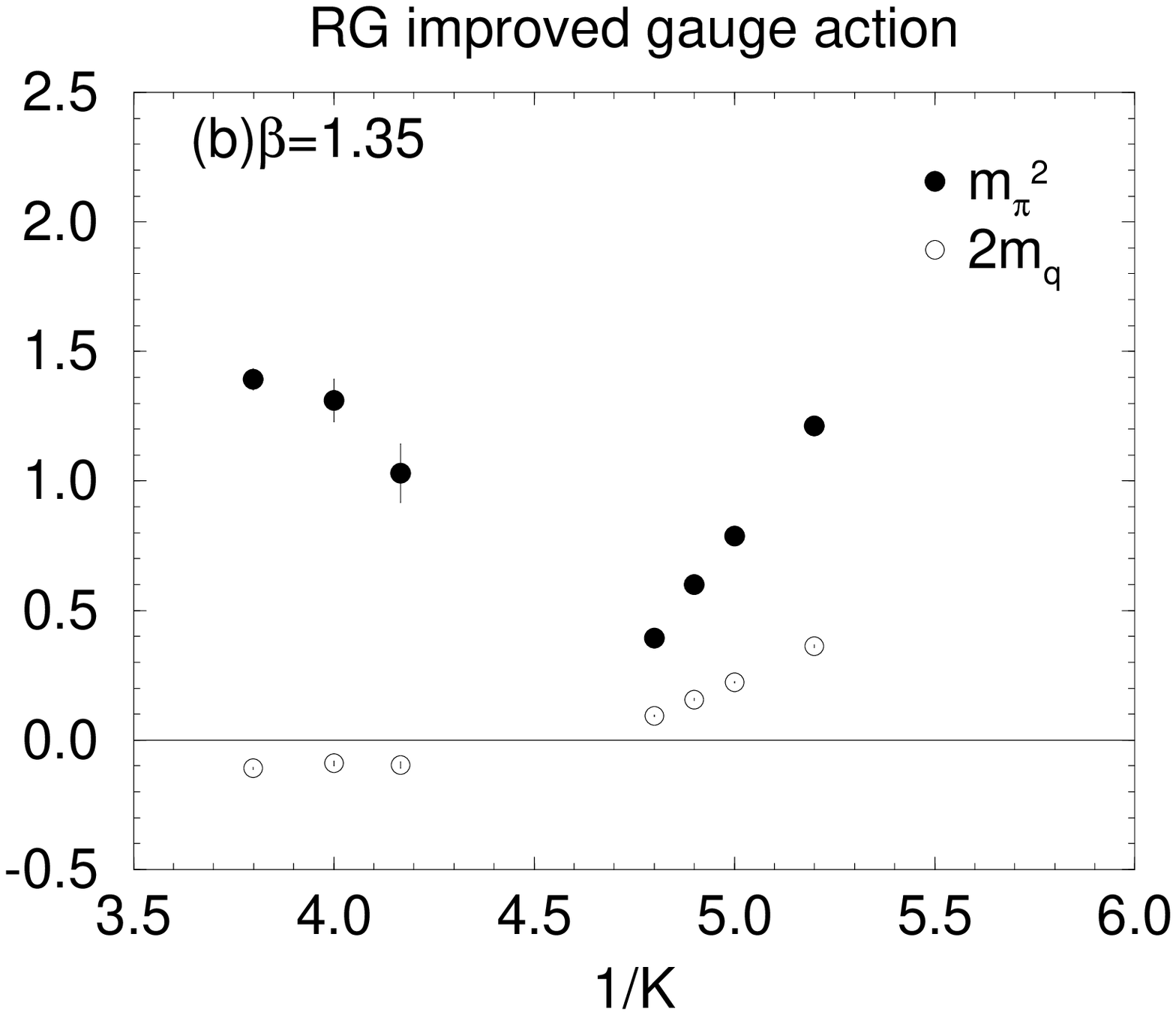}}
\vspace*{-10mm}
\caption{$\pi$ screening mass squared and quark mass as a function of $1/K$ 
for $N_f=2$ full QCD
with RG improved gauge action and Wilson quark action 
at (a) $\beta$=1.2 and (b) $\beta$=1.35,
obtained on an $8^3\times 4$ lattice periodically
doubled in one of spatial directions. }
\label{mass_RG}
\end{figure}

From our data in the previous subsection
the thermal transition around $\beta = 3.5 \sim 4.0$
seems to be a smooth cross-over at $N_T=4$, and this agrees with the prediction
that the chiral phase transition is of second order for $N_f=2$\cite{PW}
if the effect of the chiral anomaly remains at $T \approx T_c$.
However it has been reported that
the thermal transition becomes strongly first order around 
$\beta =5.0$\cite{milc}, where the thermal transition line comes very close to
the zero temperature critical line.
According to the phase diagram in the previous subsection, however,
the critical line $K_c$ defined for fixed $N_T$ does not exist
in the region where the first order transition is observed.
In fact the pion screening mass never becomes small
in the region of the first order transition.
Therefore this first order transition has nothing to do with
the chiral phase transition.
Instead it is likely that it is a lattice artifact, and
this view is indeed supported\cite{qcdpax2} 
by the fact that the strong first order
transition disappears if one uses the renormalization
group (RG) improved action for gauge fields\cite{iwasaki}.

One may wonder whether our findings such as the parity-flavor breaking phase,
the multiple structure of the critical lines, the formation of the cusp of 
the critical lines at finite temperature, and the relation of the thermal 
transition line to the critical line, are universal, and not a lattice artifact
like the first order transition at $\beta \approx 5.0$ .

To partly answer this question, we have made a new run
using the RG improved gauge action with 2 flavors of Wilson quark 
action\cite{new1}.
In Fig.\ref{mass_RG}
the pion screening mass squared and the quark mass are 
shown as a function of $1/K$
at (a) $\beta =1.2$ and (b) $\beta=1.35$ on an $8^3\times 4$ lattice,
periodically doubled in one of spatial directions for mass measurements.
These results show
that there are two critical points $K_c$ at  $\beta=1.2$,
whose gap then becomes very narrower or disappears
at $\beta = 1.35$.

Therefore at least the multiple structure of the critical lines
and the formation of the cusp of the critical line at finite temperature
are common properties in both the standard gauge action
and the RG improved one, and they may well be not
a lattice artifact. 

\begin{figure}[bt]
\centerline{\epsfxsize=65mm \epsfbox{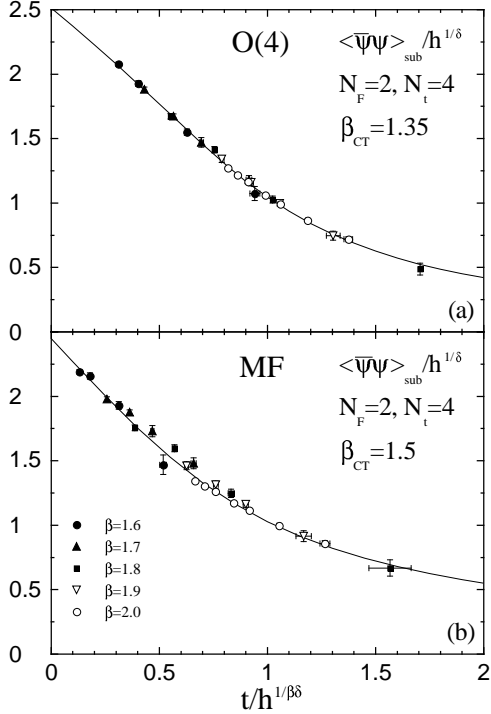}}
\vspace*{-10mm}
\caption{Best fits to the scaling function with (a) $O(4)$ and
(b) MF exponents. The plots contain all the data within the range 
$0< 2m_qa<0.8$ and $\beta \le 2.0$. Solid curves are scaling functions 
obtained in an $O(4)$ spin model\protect{\cite{scaling}} and in a
MF calculation, respectively.}
\label{exponent}
\end{figure}

\subsubsection{Scaling analysis}

Assuming the chiral phase transition 
for 2 flavor QCD is of second order,
the magnetization $M$ is expected to
be described by a single scaling function,
\begin{equation}
 M / h^{1/\delta} = f(t/h^{1/\beta\delta})
\label{scale}
\end{equation}
where $h$ is the external magnetic field and $t=[T-T_c(h=0)]/T_c(h=0)$
is the reduced temperature. For 3 dimensional $O(4)$ models, the critical
exponents are given by $1/(\beta\delta)=0.537(7)$ and $1/\delta =$ 0.2061(9)
\cite{KK}. 

Due to the explicit chiral symmetry breaking of the Wilson quark,
eq.~(\ref{scale}) contains the finite lattice spacing errors,
which may be reduced by the improved action.
Indeed it has been reported that eq.~(\ref{scale}) is
numerically satisfied for the RG improved action\cite{qcdpax2}
as follows.

Usually $M$ is identified as  $\langle \bar\psi\psi \rangle $.
For Wilson quarks, a proper subtraction 
is necessary to obtain the correct continuum limit.
The properly subtracted $\langle \bar\psi\psi \rangle$
defined via the axial Ward identity\cite{qmass} as
\begin{equation}
\langle \bar\psi\psi \rangle_{\rm sub} = 2m_qa (2K)^2 \sum_x \langle\pi(x)\pi(0)
\rangle
\end{equation}
is proposed to be used for $M$, 
together with $h= 2m_q$ and $t=\beta-\beta_{ct}$,
where $\beta_{ct}$ is a free parameter\cite{qcdpax2}. 
Assuming that the scaling function $f(x)$ is universal, 
and thus using the one obtained for an $O(4)$ model
\cite{scaling}, one can fit the data by adjusting $\beta_{ct}$ and the
scales for $t$ and $h$, with the exponents fixed to the $O(4)$ values. 
This works well with $\beta_{ct}=1.35$, as shown in Fig.~\ref{exponent}(a).
Since it is suggested that the QCD chiral transition might be
described by the mean-field (MF) criticality\cite{Kogut},
one may also test if  the MF scaling function together with 
the MF exponents, $1/(\beta\delta)=2/3$ and  $1/\delta =
1/3$, is consistent with the data.
Fig.~\ref{exponent}(b) shows, however, that it
gives no reasonable fit\cite{qcdpax2}.
The success of this scaling test with $O(4)$ exponents 
suggest not only that the chiral transition in the continuum limit
is described by the $O(4)$ criticality but also that
the chiral transition is of second order for 2 flavors.
It also shows that the RG improved gauge action indeed reduces the 
violation of the scaling relation, eq.~(\ref{scale}).


\subsection{Results for 3 and more flavors at $N_T=4$}

\begin{figure}[bt]
\centerline{\epsfxsize=65mm \epsfbox{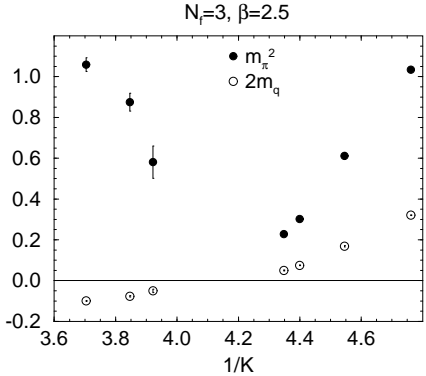}}
\vspace*{-10mm}
\caption{$\pi$ screening masses and quark mass as a function of $1/K$ for
$N_f=3$ full QCD
with the Wilson quark action 
obtained on an $8^2\times 10\times 4$ lattice periodically
doubled in a spatial direction of length 10. }
\label{mass_nf3}
\end{figure}

\begin{figure}[bt]
\centerline{\epsfxsize=65mm \epsfbox{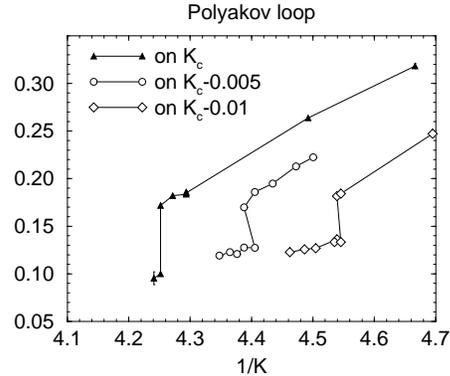}}
\vspace*{-10mm}
\caption{Polyakov loop as a function of $1/K$ along the lines
of $K=K_c(\beta)$, $K=K_c(\beta)-0.005$  and $K = K_c(\beta)-0.01$,
for $N_f=3$ full QCD on an $8^2\times 10\times 4$ lattice.}
\label{pol_nf3}
\end{figure}

The thermal chiral transition is a smooth cross-over for
2 flavors of Wilson quark action, and hopefully it will show
the criticality of the second order phase transition in
the continuum limit. On the other hand, 
the chiral transition of QCD for $N_f > 3$ is predicted to be 
of first order in the continuum\cite{PW}.
Therefore, the thermal transition with 3 or more light quarks should
also be of first order at least in the scaling region,
even though the chiral symmetry is explicitly broken by the Wilson term.

Actually a strong first-order transition has been observed for
$N_f=3$ at $\beta \approx 4.0\sim 4.7$
on an $8^2\times 10\times 4$ lattice\cite{qcdpax3}.
Assuming the behavior of the critical lines at fixed $N_T$
is similar for $N_f=2$ and 3, their data also tell us that
the tip of the cusp of the critical line, if it exists, is located around
$\beta\approx 3.0$. Like the $N_f =2$ case
the signal for the first order transition is only observed
far away from this tip of the cusp. 
Therefore, in order to claim that the thermal 
transition with three light Wilson quarks is of first order at $N_T=4$,
one should find the first order signal near the tip of the cusp around 
$\beta \approx 3.0$. 
A more fundamental question is whether
the phase structure we find for $N_f=2$,
such as the multiple structure of the critical lines
and the formation of the cusp at fixed $N_T$ 
holds also for $N_f=3$ or more.
Moreover if the thermal transition near the tip of the cusp is of first order 
as expected,
we may determine the precise location of $K_t$, so that
the question whether the thermal line $K_t$ and the cusp of the critical line 
$K_c$ touch each other or not may be clarified,
whereas it is difficult to answer it for a smooth cross-over like
in the $N_f$=2 case.

In order to answer these questions
we have made a new run for $N_f=3$ QCD using the Hybrid R algorithm\cite{new1}.
In Fig.~\ref{mass_nf3} the pion mass squared and the quark mass are again 
plotted as a function of $1/K$ at $\beta =2.5$ on an $8^2\times 10\times 4$ 
lattice. The linear extrapolation of these quantities confirms
the existence of two $K_c$'s. Like the $N_f=2$ case
the multiple structure of the critical lines presents for $N_f=3$.
Although we have not made similar measurement at $\beta =3.0$,
we expect, from the data of previous study\cite{qcdpax3},
that the critical line turns back toward the strong coupling limit,  
forming a cusp around $\beta =3.0$.

As far as the finite temperature chiral transition is concerned,
we observe that the strong first-order signal previously found around
$\beta =4.7$ becomes weaker toward the stronger coupling,
as is in the case of $N_f=2$.
To find a signal of the first-order transition near the tip of the cusp,
we have made two sequences of runs varying $K$ and $\beta$ in accordance
with relation that $K = K_c(\beta)-0.005$ or $K = K_c(\beta)-0.010$,
where $K_c(\beta)$ is the critical line defined at $T=0$, 
not at finite temperature.
Two state signals very close to the tip of the cusp, 
which is expected to exist around
$K=0.235$(: $1/K=4.255$ ) and $\beta =3.0$, are observed.
In Fig.~\ref{pol_nf3}, we plotted the behavior of the Polyakov loop
as a function of $1/K$ along the lines of $K = K_c(\beta)-0.005$ or
$K = K_c(\beta)-0.01$, together with the data previously obtained
on the line of $K_c(\beta)$\cite{qcdpax3}(, where the left most point may
not be thermalized). The signals of the first-order
transition are  seen. Therefore the thermal transition around
the tip of the cusp at $N_T=4$ may be of first order for 3 flavors, as expected
from the continuum.
It is noted, however, that this data is taken for fixed step size 
$\delta \tau =0.01$ of the Hybrid R algorithm. Therefore
more runs with a smaller $\delta\tau$ is need
to definitely claim the existence of the first order phase transition near the 
tip of the cusp.

\begin{figure}
\centerline{\epsfxsize=65mm \epsfbox{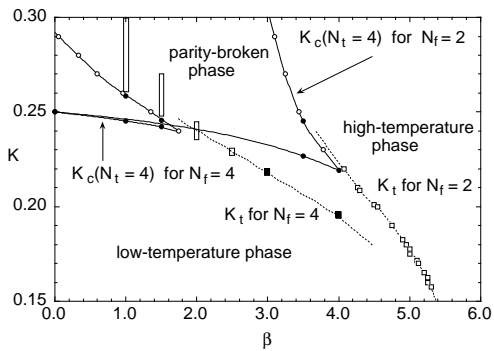}}
\vspace*{-10mm}
\caption{Phase diagram for the $N_f=4$ system for $N_t=4$.  The previous
result for $N_f=2$ is also shown for comparison.  Solid lines are
critical lines and dotted lines the line of thermal transition.  Solid
squares mark points where first-order signals are found, while
open rectangles represent region of smooth crossover.}  
\label{phase_nf4} 
\end{figure}

Now we increase the number of flavors to 4.
One of the advantage for $N_f$=4 over the $N_f=3$ case
is that the HMC method can be directly applied to it,
and thus no extrapolation of $\delta\tau$ is necessary.
Our HMC runs for $N_f=4$ with $H=0$ are carried out on an $8^3\times 4$
lattice, and hadron masses are calculated on a periodically doubled
lattice in one of spatial directions\cite{AUU2}.
The location of the critical line is estimated from 
the screening pion mass squared 
and the position and order of  the thermal
transition is examined through the behavior of physical quantities.

In Fig.~\ref{phase_nf4} we show the phase diagram for $N_f=4$
together with that for $N_f=2$ in Sec.~\ref{sec:nf2}.
We find again a cusp
structure similar to that of $N_f=2$ except for a shift of
$\delta\beta\approx 1.2$ towards stronger coupling, which  we qualitatively
expect from a larger magnitude of sea quark effects for
$N_f=4$.  We also find strong first-order signals across the thermal line
away from the tip of the cusp as marked by solid squares.  
Similar to the cases of $N_f =2$ and 3,
the transition becomes weaker towards the tip of the cusp, apparently
turning into a smooth crossover at
$\beta=2.5-2.0$.  
The location of the crossover is indicated by open rectangles
in  Fig.~\ref{phase_nf4}.

One surprising thing is that no signal of the first order transition 
can be found so far near the tip of the cusp, contrary to our expectation.
New runs are also made with an increased spatial size of $12^3\times
4$. Results do not show any deviation from those on an $8^3\times 4$ lattice.
Therefore
it is unlikely that finite-size effects
have rounded a first-order discontinuity into a smooth crossover on an 
$8^3\times 4$ lattice.  
To find the signal of the expected first order phase transition
near the tip of cusp, as in the case of $N_f=3$, simulations on the lines of 
$K=K_c(\beta)-0.005/0.01$  should be performed.
We reserve, however, this study for future investigations.

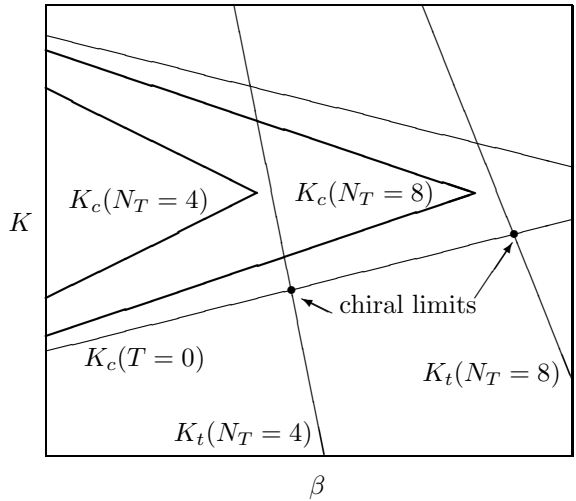
\begin{figure}[bt]

\begin{center}
\vspace*{-10mm}
\setlength{\unitlength}{1mm}
\begin{picture}(75,65)
\put(5,5){\line(1,0){70}}
\put(75,5){\line(0,1){60}}
\put(5,5){\line(0,1){60}}
\put(5,65){\line(1,0){70}}
\put(0,35){$K$}
\put(40,0){$\beta$}

\put(5,61){\line(4,-1){70}}
\put(5,19){\line(4,1){70}}

\put(5,59){\thicklines\line(3,-1){57}}
\put(5,21){\thicklines\line(3,1){57}}
\put(5,54){\thicklines\line(2,-1){28}}
\put(5,26){\thicklines\line(2,1){28}}

\put(30,65){\line(1,-5){12}}
\put(55,65){\line(2,-5){20}}

\put(37.6,27.15){\circle*{1}}
\put(67.2,34.55){\circle*{1}}

\put(22,7){$K_t(N_T=4)$}
\put(55,15){$K_t(N_T=8)$}
\put(10,17){$K_c(T=0)$}

\put(38,39){$K_c(N_T=8)$}
\put(8,38){$K_c(N_T=4)$}

\put(43,24){\vector(-2,1){4}}
\put(44,24){chiral limits }
\put(62,26){\vector(2,3){5}}

\end{picture}

\end{center}
\vspace*{-10mm}
\caption{The expected phase structure for different $N_T$'s.}
\label{phaseCL}
\end{figure}

\section{Approach toward the continuum limit} 

\begin{figure}[bt]
\centerline{\epsfxsize=65mm \epsfbox{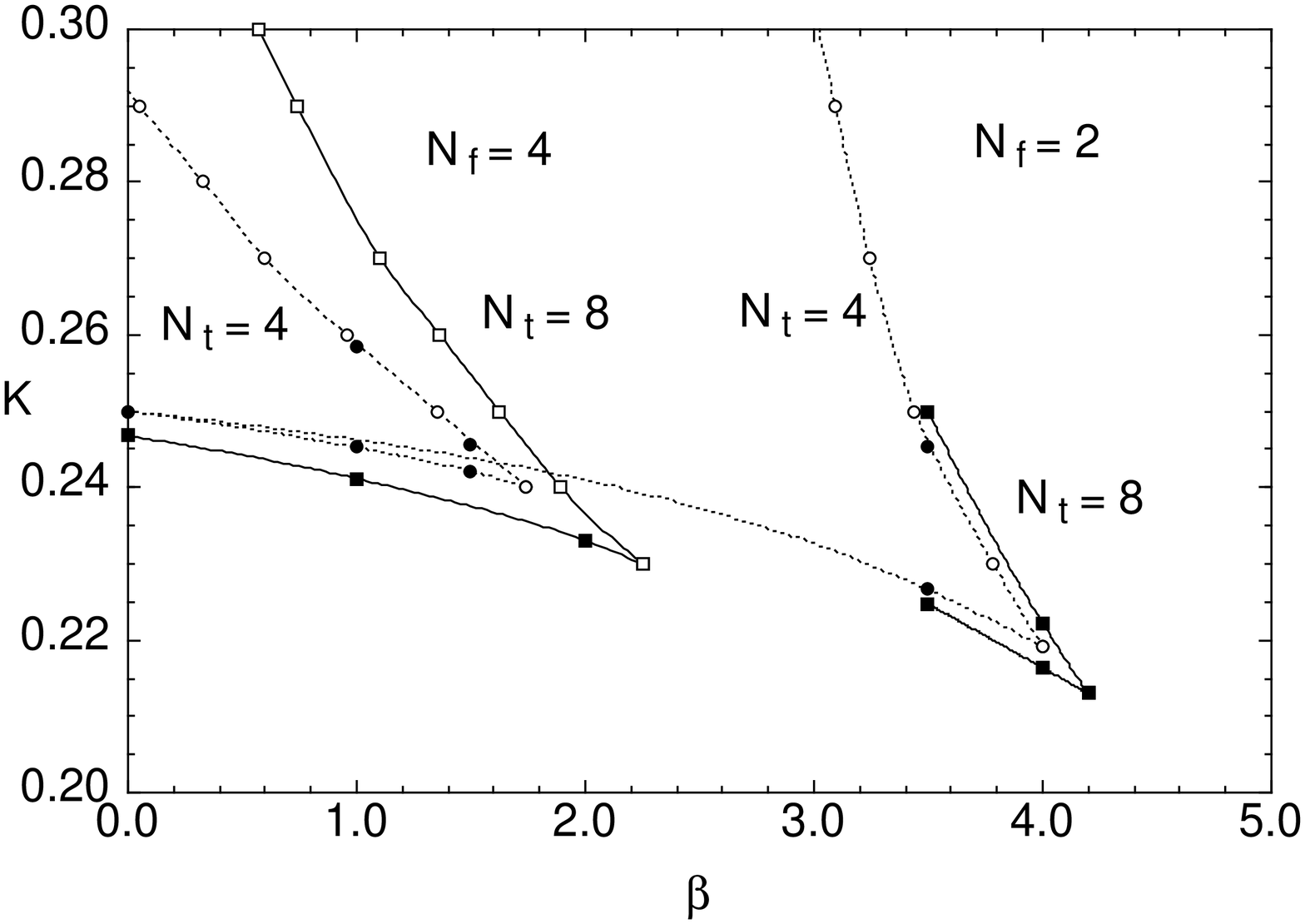}}
\vspace*{-10mm}
\caption{Location of the critical line for the lattice sizes
$8^3\times 4$ (dotted lines) and $8^3\times 8$ (solid lines) 
for $N_f=2$ and 4.}  
\label{phaseNT4-8}
\end{figure}

It is usually though that
the continuum physics of finite temperature QCD can be extracted
from the lattice simulations by the following procedure.
For fixed $N_T$ the chiral transition point may be defined
at the crossing point of $K_c$ and $K_t$, and physical observables
are measured at that point. One should repeat this for increased $N_T$,
which corresponds to the decreased lattice spacing as $a \sim 1/N_T$,
to extract the continuum values of the observables.

Based on our understanding of finite temperature phase structures
with Wilson fermion formulation, however, it now becomes clear that
the thermal transition point at fixed $N_T$
in the presence of light quarks 
can be defined anywhere in the region around the tip of the cusp.
Therefore, in principle,  no unique definition of the chiral transition point
exists at fixed $N_T$. In practice, it is better to choose
one definite prescription for the chiral transition point.
For example, as shown in Fig.~\ref{phaseCL},
the lower crossing point of $K_T(N_t)$ and $K_c(T=0)$
may be regarded as the chiral transition point, on which
observables such as $T_c/m_\rho$, etc have to be measured.
To approach the continuum limit and to investigate 
the scaling behavior, one has to increase $N_T$, repeat
same measurements on the chiral transition point corresponding to 
increased $N_T$.

One may choose freely another definition for the chiral 
limit, as the difference between two choices is $O(a)$ and should disappear in
the continuum limit. For example, the chiral limit may also be defined
at the crossing point of $K_T(N_t)$ and the line $K_q(T)$ 
where the quark mass $m_q^{WT}$ of eq.(\ref{ward}) vanishes.
Note that $K_c (T=0) = K_q (T=0)$ but
$ K_c(T\not= 0) \not= K_q(T\not= 0 )$ in general at finite temperature.
In particular $K_q( T )$ seems to run from the low temperature phase
into the high temperature phase, so that it can cross  $K_T(N_t)$. 

In order to approach the continuum limit in this way,
it is important to know how the
cusp of the critical line moves when the temporal lattice size is increased. 
To get a hint for an answer to this question,
we estimate the location of the critical lines on an
$8^3\times 8$ lattice for  $N_f=2$ and 4.

In Fig.~\ref{phaseNT4-8} we plot the critical line estimated from the pion
mass squared.  
Although the cusp is shifted toward weak coupling,
the magnitude of shift is quite small both for $N_f=2$ and 4, so that
the tip of cusp is still in the region of strong coupling for $N_t=8$.
The slow move of the tip has already been indicated
in the previous studies for $N_f=2$ \cite{qcdpax3}, whose data 
can be interpreted that the tip of the cusp is located at
$\beta=4.0-4.2$ for $N_t=6$ and at $\beta=4.5-5.0$ even for
$N_t=18$.   
If this is indeed the case,
a substantial increase in the temporal lattice size is needed 
for the tip of the cusp and the thermal transition line moving into 
the scaling region, which is usually though to start at $\beta\approx 5.5$ 
for $N_f=2$. 

It should be noted that the slow move of the tip of the cusp
has a significant impact also on spectrum calculations at zero temperature.
Since the location of the tip of the cusp 
is controlled dominantly by the smallest extension of the lattice,
the very large spatial size is necessary for an existence of the
critical hopping parameter, thus an appearance of massless pions,
in the scaling region. Otherwise the finite lattice size effect may 
become very large near the chiral limit at larger $\beta$, or
to say differently, it could happen that hadron masses are measured at some 
$\beta$ where the critical line is absent.

\section{Conclusions}

In this talk we have tried to present our understanding
on the phase structure of QCD with Wilson fermion  both at
zero and finite temperatures, based on the spontaneous breakdown
of parity-flavor symmetry.
We have shown that so far
all existing numerical data as well as analytic calculations are consistent
with our understanding.

In future study it is intersting to investigate the phase structure
of QCD with the improved Wilson fermion action (: clover action).
Since we believe that the parity-flavor breaking mechanism to obtain 
massless pions is quite general for the Wilson fermion type action 
who looses the chiral symmetry,
a similar phase structure is expected to exist.
Furthermore it is important to see how fast the tip of cusp, if it exists, 
moves
toward the scaling region as $N_T$ increases. Hopefully the improvement
of the fermion action as well as that for the gauge action may
overcome the problem of the slow move in the previous section.

Although a lot of things concerning the parity-flavor breaking mechanism
still remain to be investigated,
we think that no conceptual difficulty in Wilson-type fermion actions
at finite temperature exists.
So now it is a good time to start serious study, 
like the Kogut-Susskind quark action, to get real physical
results from the lattice QCD with Wilson-type fermion action,
but it may be necessary to use gauge and fermion 
improvements\cite{iwasaki,qcdpax2,milcimp}.

\section*{Acknowledgements}

Numerical calculations for the present work have been carried out
at Center for Computational Physics
and on VPP500/30 at Science Information Center,
both at University of Tsukuba. 
I thank my collaborators, Y. Iwasaki, K. Kanaya, T. Kaneda, S. Kaya, 
A. Ukawa, and T. Yoshi\'e for useful discussions.
I also thank Dr. G. Boyd for useful comments.
This work is supported in part by the Grants-in-Aid of 
the Ministry of Education (Nos. 08640350, 09246206).


\begin{thebibliography}{9}

\bibitem{earlywork}M. Fukugita, S. Ohta and A. Ukawa,
Phys. Rev. Lett. 57 (1986) 1974; A. Ukawa, Nucl. Phys. B(Proc.
Suppl.)9 (1990) 463. See also R. Gupta {\it et
al.,} Phys. Rev. D40 (1989) 2072; K. M. Bitar  {\it et al.,}
Phys. Rev. D43 (1991) 2396.

\bibitem{qmass}
M. Bochicchio {\it et al.}, Nucl. Phys. B262 (1985) 331;
S. Itoh {\it et al.}, Nucl. Phys. B274 (1986) 33.

\bibitem{aoki1}S. Aoki, Phys. Rev. D30 (1984) 2653; 
33 (1986) 2377; 34 (1986) 3170;
Phys. Rev. Lett. 57 (1986) 3136; Nucl. Phys. B314
(1989) 79.

\bibitem{aoki2} S.Aoki and A. Gocksch,
Phys. Lett. B231 (1989) 449; 243 (1990) 1092;
Phys. Rev. D45 (1992) 3845.

\bibitem{aoki3} S. Aoki, Prog. Theor. Phys. 122 (1996) 179.

\bibitem{EN} T. Eguchi and R. Nakayama, Phys. Lett. 126B (1983)
 89.

\bibitem{U1}S. Itoh, Y. Iwasaki and T. Yoshi\'e,
Phys. Rev. D36 (1987) 527.

\bibitem{SDB}R. Setdoodeh, C.T.H. Davies, I.M. Barbour, Phys. Lett.
B{\bf 213} (1988) 195.

\bibitem{BLLS}I.M. Barbour, E. Laermann, Th. Lippert, K. Schilling,
Phys. Rev. D{\bf 46} (1992) 3618.

\bibitem{milc}C. Bernard {\it et al.}, Phys. Rev. D46 (1992) 4741,
49 (1994) 3574, 50 (1994) 3377.

\bibitem{qcdpax1}Y. Iwasaki {\it et al.}, Phys. Rev. Lett. 67 (1991) 1494,  
69 (1992) 21; Nucl. Phys. B(Proc. Suppl.)30 (1993) 327, 
34 (1994) 314.

\bibitem{AUU1}S. Aoki, A. Ukawa and T. Umemura, Phys. Rev. Lett. 76 (1996) 873;
Nucl. Phys. B(Proc. Suppl.)47 (1996) 511.

\bibitem{AUU2}S. Aoki, T. Kaneda, A. Ukawa and T. Umemura,
Nucl. Phys. B(Proc. Suppl.)53 (1997) 438.

\bibitem{PW}R. Pisarski and F. Wilczek, Phys. Rev. D29 (1984) 338.

\bibitem{qcdpax2} Y. Iwasaki, K. Kanaya, S. Sakai and T. Yoshi\'e,
Nucl. Phys. B(Proc. Suppl.)42 (1995) 502;
Y. Iwasaki, K. Kanaya, S. Kaya, S. Sakai and T. Yoshi\'e,
Nucl. Phys. B(Proc. Suppl.)47 (1996) 515;
Y. Iwasaki, K. Kanaya, S. Kaya and T. Yoshi\'e,
Phys. Rev. Lett. 78 (1997) 179.

\bibitem{iwasaki} Y. Iwasaki, Nucl. Phys. B258 (1985) 141;
Univ. of Tsukuba report UTHEP-118(1983), unpublished.

\bibitem{new1} S. Aoki, Y. Iwasaki, K. Kanaya, S. Kaya A. Ukawa and 
T. Yoshi\'e,
in preparation.

\bibitem{KK} K. Kanaya and S. Kaya, Phys. Rev. D51 (1995) 2404.

\bibitem{scaling} D. Toussaint, Phys. Rev. D55 (1997) 362.

\bibitem{Kogut} A. Koci\'c and J. Kogut, Phys. Rev. Lett. 74 (1995) 3108.

\bibitem{qcdpax3} Y. Iwasaki, K. Kanaya, S. Sakai and T. Yoshi\'e,
Z. Phys. C71 (1996) 337;
Y. Iwasaki, K. Kanaya, S. Kaya, S. Sakai and T. Yoshi\'e,
Nucl. Phys. B(Proc. Suppl.)42 (1995) 499;
Z. Phys. C71 (1996) 343; Phys. Rev. D54 (1996) 7010.

\bibitem{milcimp}C. Bernard {\it et al.}, 
Nucl. Phys. B(Proc. Suppl.)53 (1997) 446.
\end{thebibliography}
\end{document}